\documentclass[fleqn,twoside]{article}
\usepackage{graphicx,epsfig,espcrc2}

\newcommand{\AmS}{{\protect\the\textfont2
  A\kern-.1667em\lower.5ex\hbox{M}\kern-.125emS}}
\title{Search for aligned events in muon groups measured by BUST}%

\author{A.L.Tsyabuk, R.A.Mukhamedshin and Yu.V.Stenkin
\address{Institute for Nuclear Research of Russian Academy of Sciences, \\
        60th October anniv. prosp., 7a, Moscow 117312, RUSSIA,\\
         e-mail: stenkin@sci.lebedev.ru}}%

\begin{document}

\begin{abstract}
A search for aligned events has been done throughout the muon
groups events measured by Baksan Underground Scintillation
Telescope (BUST) during a period of 7.7 years. Only groups of
multiplicity $>3$ for muon threshold energy equal to 0.85 TeV were
selected for the analysis. A distribution of the events on
alignment parameter $\lambda$ has been obtained and compared with
the results of Monte-Carlo simulation made for this experiment.
The upper limit for aligned muon event flux as low as
$5.3\cdot10^{-15}$ $cm^{-2} sec^{-1} sr^{-1}$ is given.
\end{abstract}

\maketitle

\section{Introduction}

The muon bundle data cover an energy range between the data of
direct methods (balloons, satellites) and the Extensive Air
Showers (EAS) data: from $\sim10^{13}$ and up to $\sim10^{17}$ eV
for primary cosmic ray (i.e. includes the "knee"). High energy
muons are produced in the highest energy part of the EAS cascade
at high altitudes. The muon bundle events give information on the
transverse momenta of the secondaries in high energy collisions,
which affects a lateral spread of muons in the bundles. On the
other hand, the "knee" energy range is related to interesting
phenomena observed in cosmic rays, in particular, to the
phenomenon of the alignment of most energetic subcores of
gamma-ray-hadron ($\gamma - h$) families (particles of highest
energies in the central EAS core) found in the "Pamir" emulsion
chamber experiment \cite{pam,kap}. In this work we performed a
search for aligned events among the muon groups observed in BUST.
The first search for aligned muon bundle events has been done with
MACRO detector \cite{siol} without certain results.

\section{The experiment}

The BUST detector \cite{ale} is located in Baksan Valley (North
Caucuses, Russia) at a height of 1700 m. a. s. l. in the
underground laboratory at a distance of 550 m from the tunnel
entrance. Its effective depth is 850 $hg/cm^{2}$, and the
effective threshold energy is 220 GeV. The depth varies from ~800
$hg/cm^{2}$ for nearly vertical directions up to ~6000 $hg/cm^{2}$ for
slant trajectories where the energy threshold is about 6 - 10 TeV.
The telescope looks like a four-storeyed building with a size of
$16.7\times16.7\times11.1$ $m^{3}$. Four vertical scintillator
layers and four horizontal scintillator planes are formed by
liquid scintillation detectors of standard type. The total number
of detectors is 3150. The standard detector
($0.7\times0.7\times0.3$ $m^{3}$) consists of an aluminum tank
filled with liquid scintillator viewed by a 15-cm diameter PMT
(FEU-49).

In this experiment additional requirements were applied to select
the so called {\it aligned} events. We used the standard $\lambda$
parameter introduced by the "Pamir" experiment \cite{pam}, which is
varying from $\sim-0.5$ to $\sim1$ (for aligned events). It is equal
to near zero values for isotropic events.  Inclined tracks with
zenith angles inside a
range of 50-$70^{\circ}$ were selected to make the muon energy
threshold higher. The effective depth for such a selection was
equal to 2500 $hg/cm^{2}$, which corresponds to muon threshold
energy of 0.85 TeV.

\section{Experimental results}

\begin{figure*}[tbh]
 \begin{center}
\includegraphics[width=14.cm]{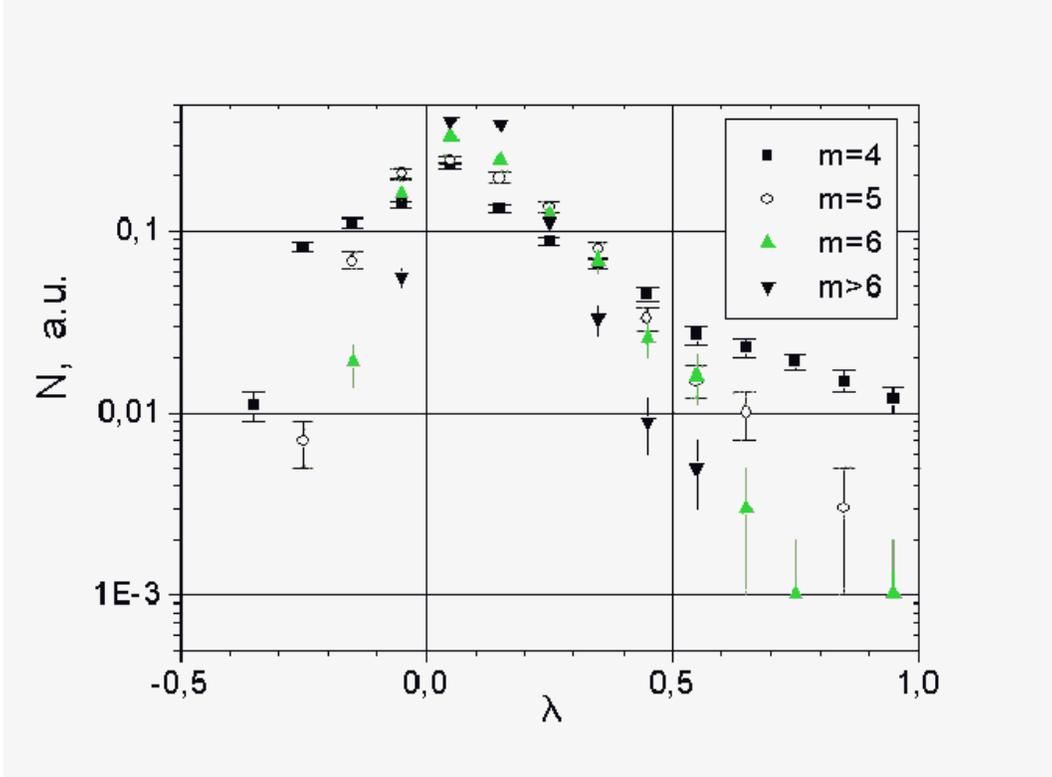}
\caption{Experimental distribution on $\lambda$ parameter.}
 \end{center}
 \end{figure*}
The experimental normalized distributions on the $\lambda$
parameter is shown in Fig.1 for visible muon multiplicity m=4, 5
and 6. As one can see we have no any significant access near
$\lambda=1$. In fact, aligned events do exist as it is seen in
Fig.1, but, very similar spot configurations can be realized by
chance as it is shown below.
\begin{figure*}[tbh]
\begin{center}
\begin{minipage} {0.475\linewidth}
\includegraphics [width=7.6cm] {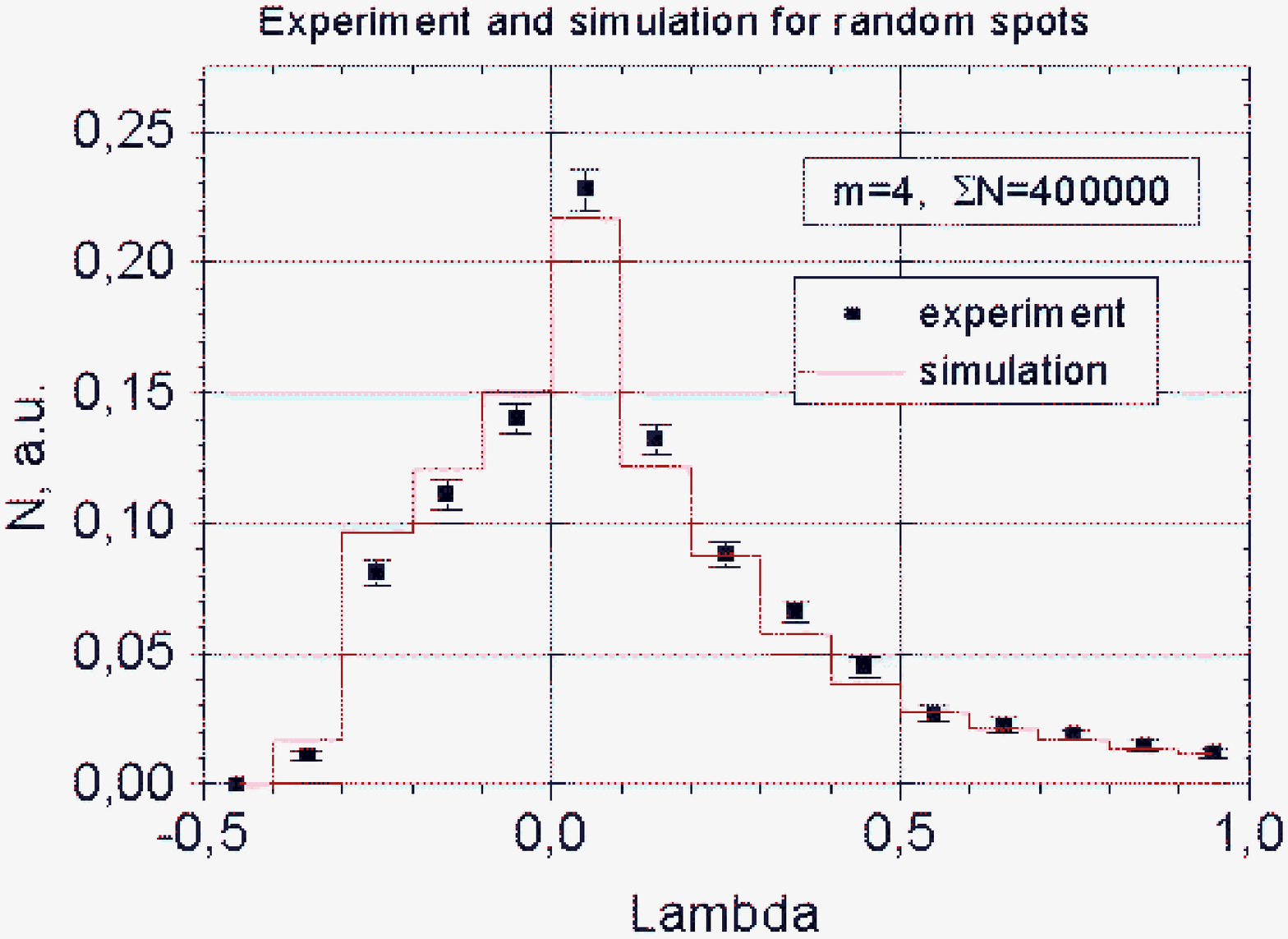}\\
 \vspace{-0.5cm}  a)
\end{minipage}
\hspace*{0.3cm}
\begin{minipage} {0.475\linewidth}
\includegraphics [width=7.6cm] {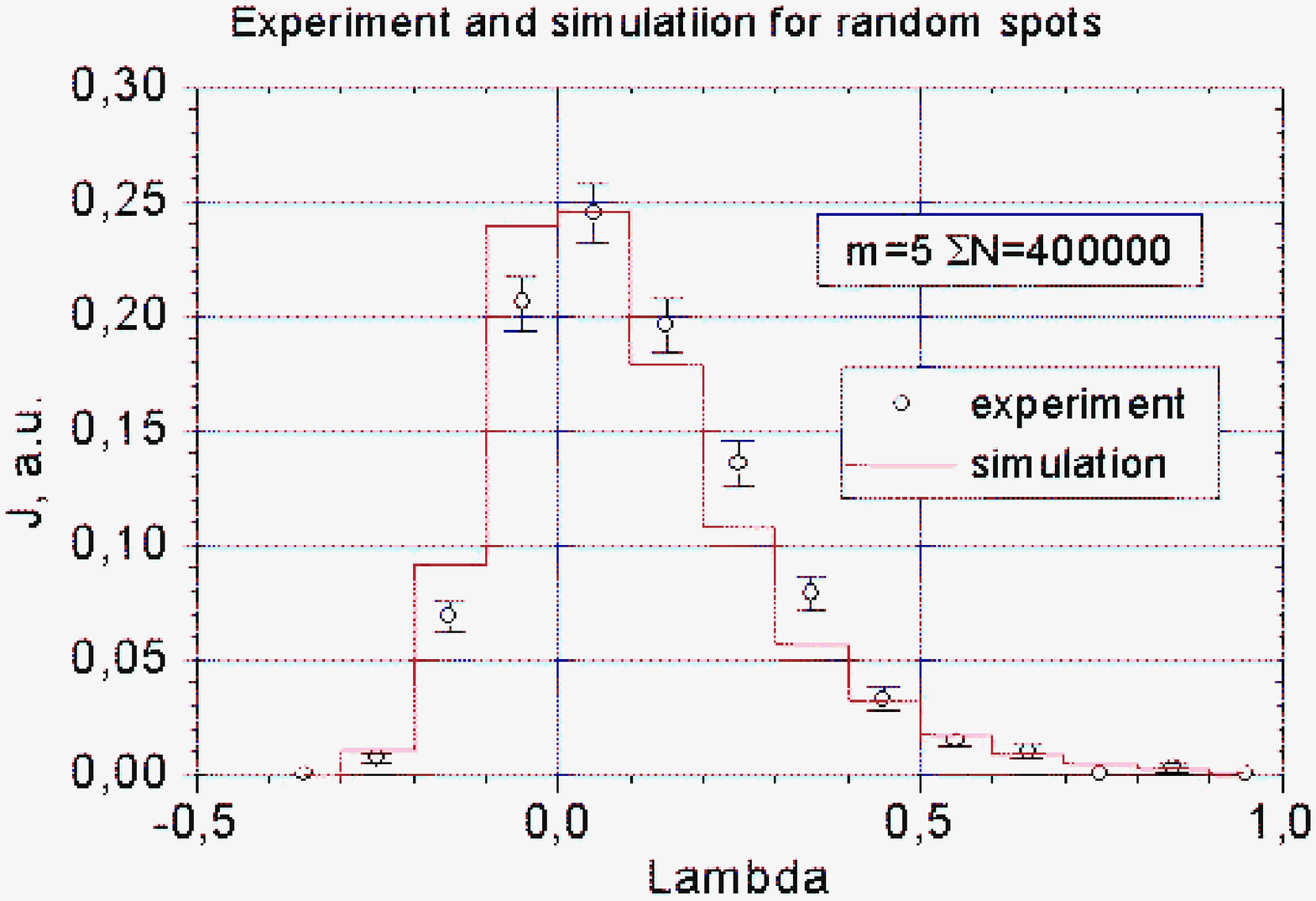}\\
 \vspace{-0.5cm}  b)
\end{minipage}
\vspace{0.5cm}
\begin{minipage} {0.475\linewidth}
\includegraphics [width=7.6cm] {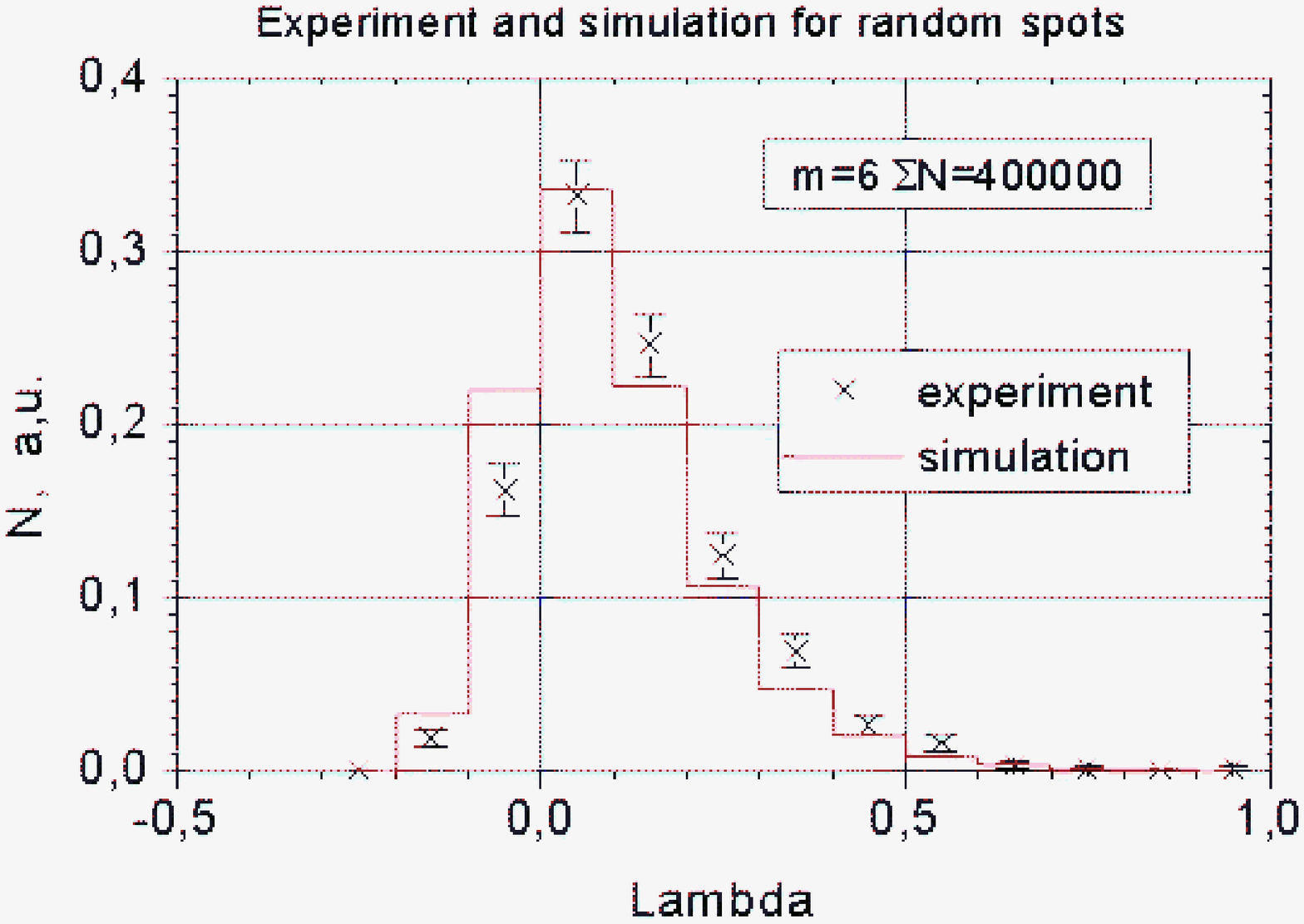}\\
 \vspace{-0.2cm} c)
\end{minipage}
\hspace*{0.3cm}
\begin{minipage} {0.475\linewidth}
\vspace{-1.5cm} \caption{Comparison of the experimental
$\lambda$-distributions with random spots  simulations for m=4
(a), m=5 (b) and m=6 (c).}
\end{minipage}
 \end{center}
\vspace{-.5cm}
\end{figure*}

\section{Simulations}

A Monte Carlo simulation of EAS propagation in the atmosphere has
been carried out. Muon groups were first simulated in the
framework of the MC0 code \cite{fed} (close to QGSJET), which
reproduces well the results obtained by the PAMIR experiment
emulsion chamber experiments and does not include unusual
processes. To analyze the alignment phenomena, a simplified model
of particle coplanar generation (PCGM) in the first interaction
of a primary particle with an air nucleus has been used. This
model reproduces PAMIR's data on aligned $\gamma - h$ families.
\begin{figure*}[tbh]
\begin{minipage} {0.475\linewidth}
\includegraphics [width=7.5cm] {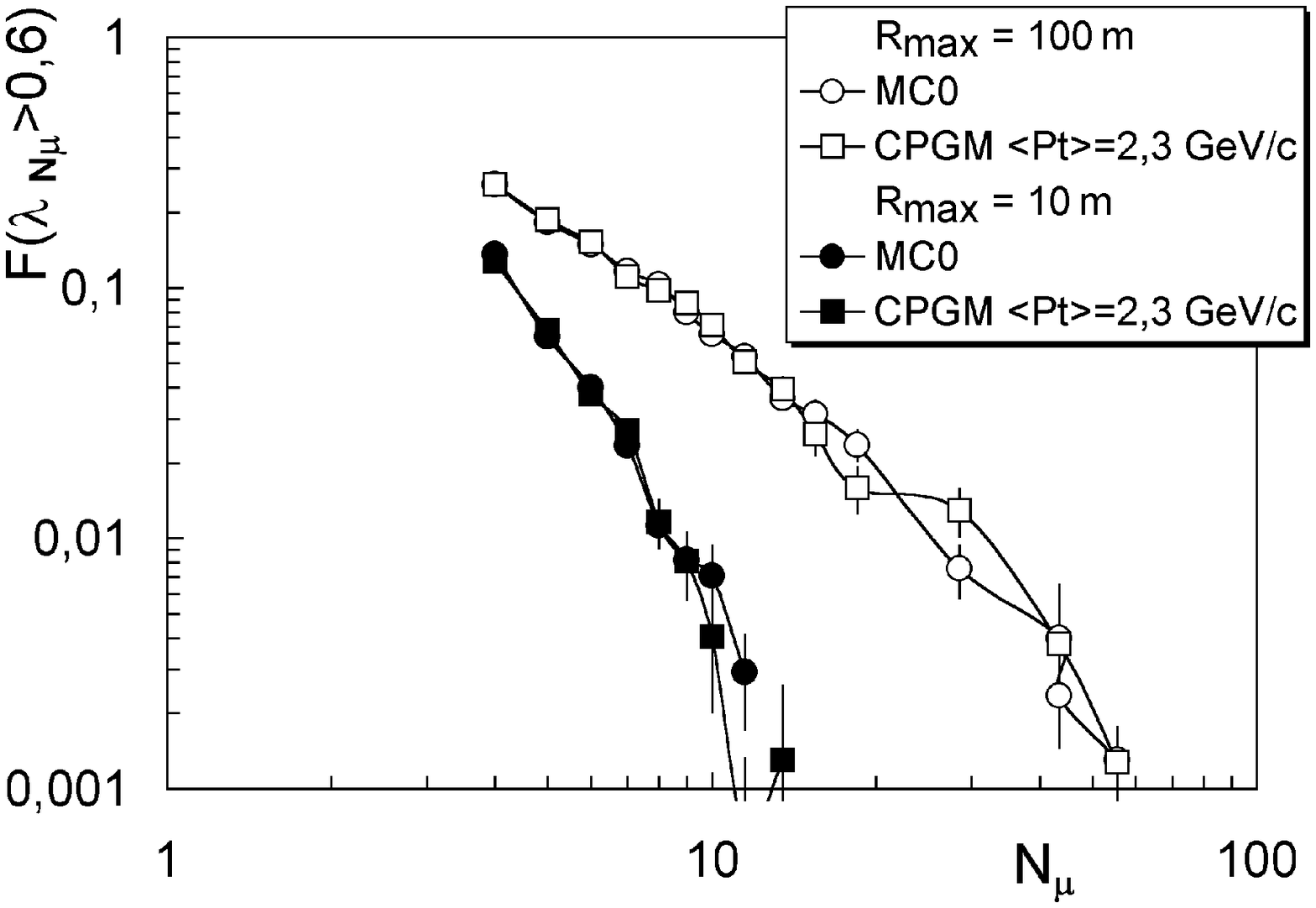}
\caption{Dependence of the fraction of aligned muon groups
 $F(\lambda_{N_{\mu}}\ge 0.6)$ simulated in the
framework of the MC0 and PCGM algorithms on muon multiplicity for
different set-up cut-off radius ($R_{max}$).}
\end{minipage}
\hspace*{0.2cm}
\begin{minipage} {0.475\linewidth}
\includegraphics [width=7.5cm] {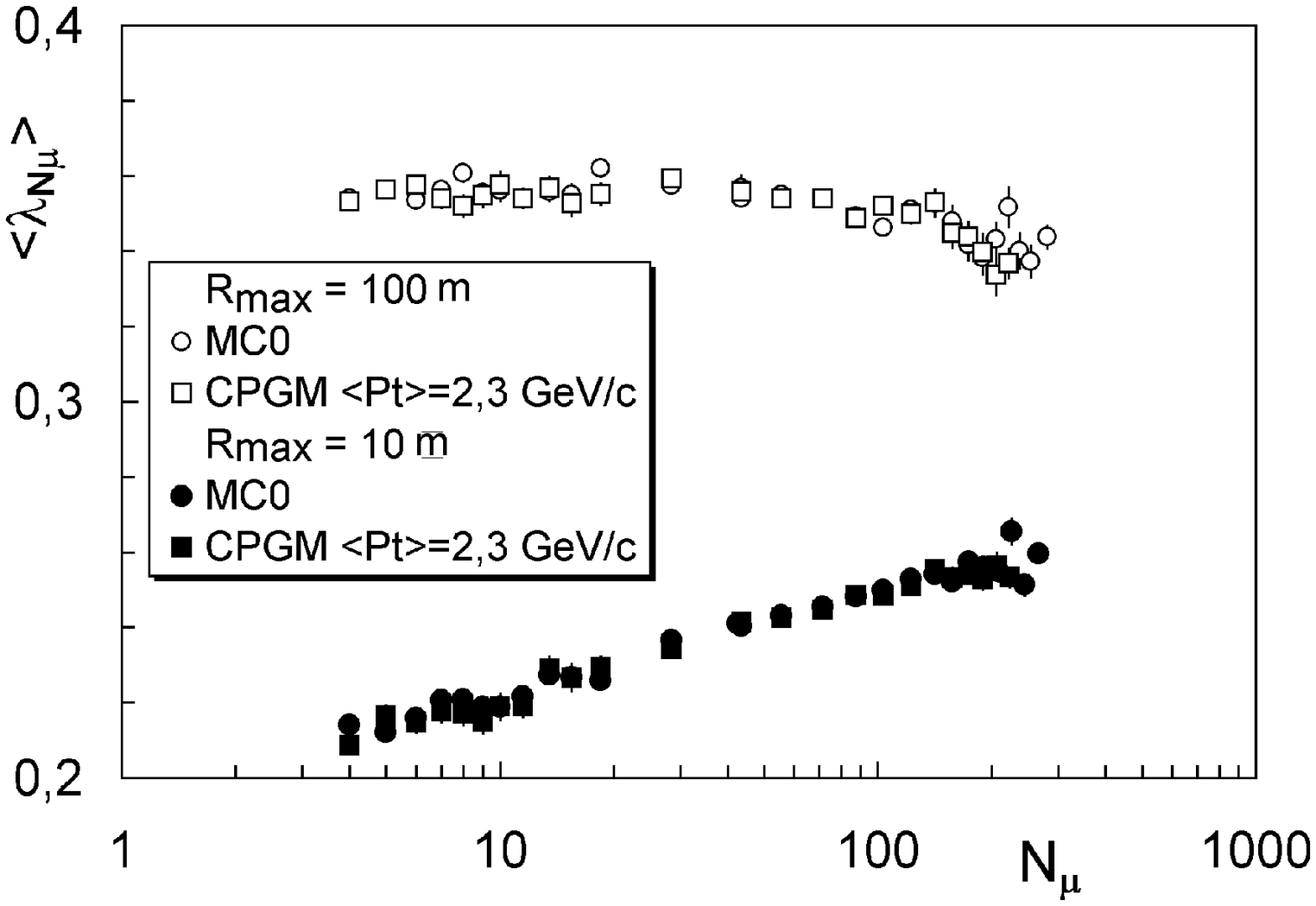}
\caption{Dependence of the average alignment $<\lambda_{N_{m}}>$
simulated in the framework of the MC0 and PCGM algorithms on muon
multiplicity for different set-up cut-off radius (Rmax).}
\end{minipage}
 \end{figure*}
%
Conclusions found on the
basis of the simulations are as follows: \\
1. The influence of magnetic field on alignment of muons is strong
and it doubles the background fraction of aligned muon bundles
caused by the cascade development.\\
2. In accordance with the PAMIR's concept assuming the coplanar
generation of several most energetic particles, which cannot
actually decay into muons, the appearance of coplanar generation
has no influence on the features of muon groups formed mainly by
muons produced by lower-energy secondary particles in further
generations. Figure 3 demonstrates this: the results of simulations
made in the
framework of both the MC0 and PCGM algorithms coincide for all
the muon multiplicities while there exists a dependence on the
detector dimensions. For $\sim10-m$ set-ups (BUST dimensions) it
is close to that being characteristic for $\gamma - h$ families.
As one can see, a fraction of aligned events on a level of about
several percent is normal and this is in agreement with our
experimental data shown in Fig. 1. Fig.4 shows the same behaviour
for the average alignment $<\lambda_{N_{\mu}}>$.\\
3. A significant alignment of muons could only be produced in
case of direct coplanar muon generation.

\section{Summary}

 An experimental search has been performed for aligned events
in muon groups with threshold energy equal to 0.85 TeV. The
measured distributions on the alignment parameter $\lambda$ agree
very well with simulated event distributions and there is no
evidence of the existence of such events among the muon groups
underground (at least for threshold energy 0.85 TeV) at a level
beyond the expectations obtained for simulated events with a
random track distribution. Moreover, even in EAS's simulated in
the framework of a model including the coplanar particle
generation there exists no visible effect of alignment in muon
groups because only a small fraction of muons could be generated
in the first interaction via the decay of the most energetic
aligned secondaries. Fluctuations of cascade development and a
huge number of muons produced in later generations make this
process random. Therefore, the observed muon alignment is only
determined by random fluctuations of muon tracks on the detector
area. Taking into account the detector area ($\sim200$ $m^{2}$ )
and the trigger selected solid angle ($\sim0.38$ sr) and the duration
of the experiment (7.7 y) we can put an upper limit to the flux of
aligned muon groups on a one sigma level as low as $F_{aligned}<
5.3\cdot10^{-15}$ $cm^{-2}$ $sec^{-1}$ $sr^{-1}$.

This work was supported in part by the RFBR grants, projects No.
04-02-17083a, 03-02-16272a, and 03-02-17465a; and Ministry of
Education and Science, projects SSL-1828.2003.02 and
SSL-1782.2003.02.

\end{document}